\documentclass[screen, review, sigconf]{acmart}
\usepackage[utf8]{inputenc}
\usepackage[T1]{fontenc}
\usepackage{graphicx}
\usepackage[english]{babel}
\usepackage{amsmath}
\usepackage{caption}
\usepackage{subcaption}
\usepackage{balance}
\usepackage{wrapfig}
\usepackage{hyperref}
\usepackage{multirow}
\usepackage{multicol}
\usepackage{color}
\usepackage{xcolor}
\usepackage{colortbl}
\usepackage{booktabs}
\usepackage{flushend}
\usepackage{todonotes}

\usepackage{pifont}

\newsavebox{\measurebox}

\setcopyright{open}
\acmConference[KiML'20]{KDD Workshop on Knowledge-infused Mining and Learning}{August 24, 2020, San Diego, California, USA}

\author{Amanuel Alambo}
\affiliation{%
 \institution{Knoesis Center}
  \city{Dayton, Ohio}
}
\email{amanuel@knoesis.org}

\author{Manas Gaur}
\affiliation{%
  \institution{AI Institute, University of South Carolina}
  \city{Columbia, South Carolina}
}
\email{mgaur@email.sc.edu}

\author{Krishnaprasad Thirunarayan}
\affiliation{%
  \institution{Knoesis Center}
  \city{Dayton, Ohio}
}
\email{tkprasad@knoesis.org}

\begin{document}

\title{
Depressive, Drug Abusive, or Informative: \\
Knowledge-aware Study of News Exposure during COVID-19 Outbreak
}

\begin{abstract}
The COVID-19 pandemic is having a serious adverse impact on the lives of people across the world. COVID-19 has exacerbated community-wide depression, and has led to increased drug abuse brought about by isolation of individuals as a result of lockdown. Further, apart from providing informative content to the public, the incessant media coverage of COVID-19 crisis in terms of news broadcasts, published articles and sharing of information on social media have had the undesired snowballing effect on stress levels (further elevating depression and drug use) due to uncertain future. In this position paper, we propose a novel framework for assessing the spatio-temporal-thematic progression of depression, drug abuse, and informativeness of the underlying news content across the different states in the United States. Our framework employs an attention-based transfer learning technique to apply knowledge learned on a social media domain to a target domain of media exposure. To extract news articles that are related to COVID-19 communications from the streaming news content on the web, we use neural semantic parsing, and background knowledge bases in a sequence of steps called semantic filtering. We achieve promising preliminary results on three variations of Bidirectional Encoder Representations from Transformers (BERT) model. We compare our findings against a report from Mental Health America and the results show that our fine-tuned BERT models perform better than vanilla BERT. Our study can benefit epidemiologists by offering actionable insights on COVID-19 and its regional impact. Further, our solution can be integrated into end-user applications to tailor news  for users  based on their emotional tone measured on the scale of depressiveness, drug abusiveness, and informativeness.
\end{abstract}


\maketitle
\pagestyle{plain}

\section{Introduction}
\label{sec:I}
COVID-19 pandemic has changed our societal dynamics in different ways due to the varying impact of the news articles and broadcasts on a diverse population in the society. Thus, it is important to place the news articles in their spatio-temporal-thematic (\citeauthor{nagarajan2009spatio}, \citeyear{nagarajan2009spatio}; \citeauthor{andrienko2013thematic}, \citeyear{andrienko2013thematic}; \citeauthor{harbelot2015lc3}, \citeyear{harbelot2015lc3}) contexts to offer appropriate and timely response and intervention.  In order to limit the scope of this research agenda, we propose to focus on identifying regions that are exposed to depressive and drug abusive news articles and to determine/recommend ways for timely interventions by epidemiologists. 

The impact of COVID-19 on mental health has been investigated in recent studies (\citeauthor{garfin2020novel}, \citeyear{garfin2020novel}; \citeauthor{holmes2020multidisciplinary}, \citeyear{holmes2020multidisciplinary}; \citeauthor{qiu2020nationwide}, \citeyear{qiu2020nationwide}). \cite{garfin2020novel} studied the impact of repeated media exposure on the mental well-being of individuals and its ripple effects. \cite{holmes2020multidisciplinary} underscore the importance of a multidisciplinary study to better understand COVID-19. Specifically, the study explores its psychological, social, and neuroscientific impacts. \cite{qiu2020nationwide} studied the psychological impact COVID-19 lockdown had on the Chinese population. These studies, however, do not adequately explore a technique to computationally analyze the regional repercussions associated with media exposure to COVID-19 that may provide a better basis for local grassroots level action. 

We propose an approach to measure depressiveness, drug abusiveness, and informativeness as a result of media exposure for various states in the US in the months from January 2020 to March 2020. Our study is focused on the first quarter of 2020 as this period was critical in the spread of COVID-19 and its ominous impact; this was a period when the public faced major changes to lifestyle including lockdown, social distancing, closure of businesses, unemployment, and broadly speaking, complete lack of control over the unfolding situation precipitating in severe uncertainty about the impending future. In consequence, this continued media exposure progressively worsened the mental health of individuals across the board. We analyze and score news content on three orthogonal dimensions: spatial, temporal, and thematic. For spatial, we use state boundaries. For temporal, we use monthly data analysis. For thematic, we score news content on the category/dimension of depression, drug abuse and informativeness (relevant to COVID-19 but not directly connected to either depression or drug-abuse).

\begin{figure}[htbp]
\centerline{\includegraphics[width=70mm]{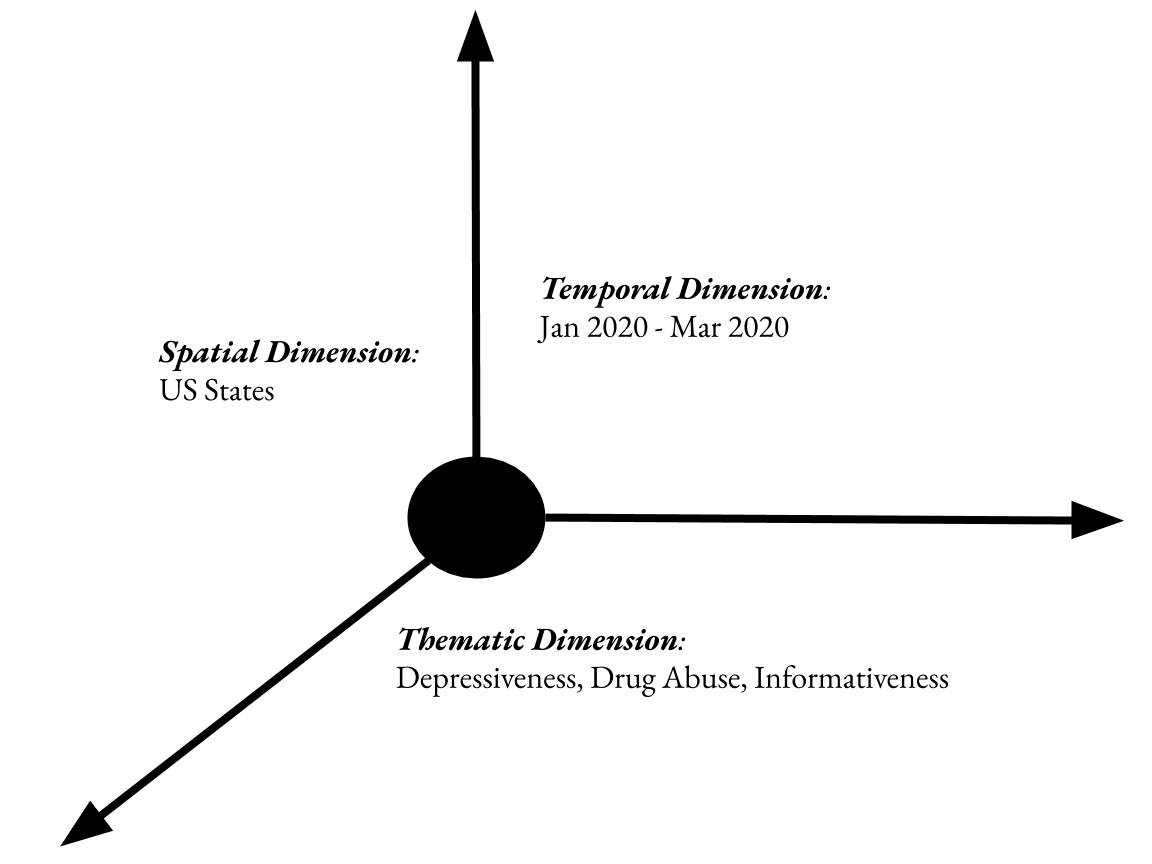}}

\vspace*{-0.3cm}
\caption{Spatio-Temporal-Thematic Dimensions}
\label{fig}
\end{figure}

Our study hinges on the use of domain-specific language modeling and transfer learning to better understand how depressiveness, drug abusiveness, and informativeness of news articles evolve in response to media exposure by people. We conduct the transfer of knowledge learned on a social media platform to the domain of exposure to news using variations of the attention-based BERT model (\citeauthor{devlin2018bert}, \citeyear{devlin2018bert}), also called Vanilla BERT. Thus, in addition to vanilla BERT, we fine-tune BERT models on corpora that are representative of depression and drug abuse. Then, we compare results obtained using the three variants of the BERT model. For scoring depressiveness, drug abusiveness, and informativeness of news articles, we utilize entities from structured domain knowledge from the Patient Health Questionnaire (PHQ-9) lexicon (\citeauthor{yazdavar2017semi}, \citeyear{yazdavar2017semi}), Drug Abuse Ontology (DAO) (\citeauthor{cameron2013predose}, \citeyear{cameron2013predose}), and DBpedia (\citeauthor{lehmann2015dbpedia}, \citeyear{lehmann2015dbpedia}). PHQ-9 lexicon is a knowledge base developed specifically for assessing depression, and DAO is built to study drug abuse. Similarly, we use DBpedia, which is a generic and comprehensive knowledge base, for assessing the informativeness of news content.

Having determined the scores for depressiveness, drug abusiveness, and informativeness of news articles for each state during the three months, we computed the aggregate score for each thematic category by summing up the scores for the news articles. We finally assigned the category with the highest score as a label for a state. For instance, if the aggregate score of depressiveness for the state of Iowa in the month of January 2020 is the highest of the three thematic categories, then the state of Iowa is assigned a label of depression for that month, which means the state of Iowa is most exposed to depressive news contents. Thus, identifying which states are consistently exposed to depressive or drug abusive news contents enables policy makers and epidemiologists to devise appropriate intervention strategies.

\section{Data Collection}
\label{sec:EDA}

We collected 1.2 Million news articles from the Web and GDELT\footnote{\url{https://www.gdeltproject.org/}} (a resource that stores world news on significant events from different countries) using semantic filtering (\citeauthor{sheth2016semantic}, \citeyear{sheth2016semantic}; \citeauthor{arachie2020unsupervised}, \citeyear{arachie2020unsupervised}; \citeauthor{gaur2020semantics}, \citeyear{gaur2020semantics}) and spanning the period from January 01, 2020, to March 29, 2020. We filtered news articles that did not originate from within the US and grouped the ones that are from the US based on their state of origination. The state-level grouped news articles had a total of over 150K entities identified using DBpedia spotlight service\footnote{\url{https://www.dbpedia-spotlight.org/}}. However, since using a coarse filtering service such as DBpedia spotlight over the entire news articles is not efficient and brings in irrelevant entities, and thus noisy news articles, we utilize (``$\mbox{i}$'') a neural parsing approach with self-attention (\citeauthor{wu2019npa}, \citeyear{wu2019npa}) to extract relevant entities. After extracting relevant entities and news articles, we use (``$\mbox{ii}$'') DBpedia spotlight service to identify news articles that are related to online communications about COVID-19.

\begin{figure}[htbp]
\centerline{\includegraphics[width=80mm]{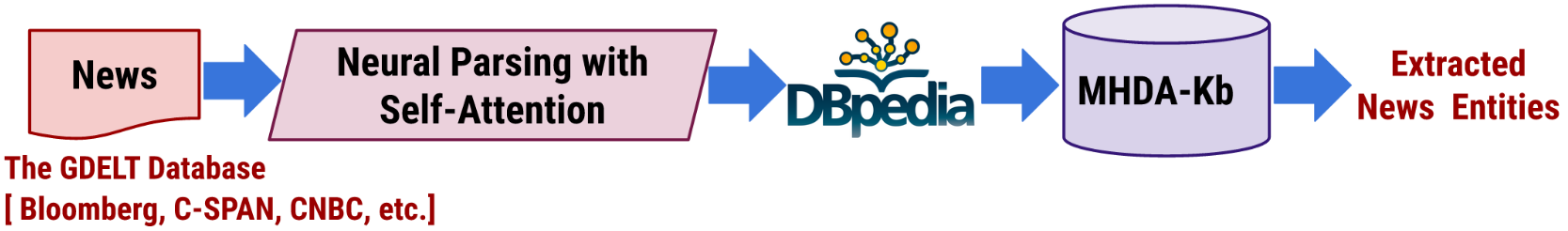}}

\vspace*{-0.3cm}
\caption{Knowledge-based entity extraction using Semantic Filtering}
\label{fig}
\end{figure}

For this task, we explored 780 DBpedia categories that are relevant to COVID-19 communications to create the most relevant set of entities and news articles. Further, upon inspection of the news articles, we discovered medical terms that were not available in DBpedia. As a result, we used (``$\mbox{iii}$'') the MeSH terms hierarchy in Unified Medical Language System (UMLS), the Diagnostic and Statistical Manual for Mental Disorders (DSM-5) lexicon (\citeauthor{gaur2018let}, \citeyear{gaur2018let}), and Drug Abuse Ontology (DAO), collectively referred to as \textit{Mental Health and Drug Abuse Knowledgebase} (MHDA-Kb) to spot additional entities. Thus, from ~700K unique news articles (which are extracted from the total of 1.2 Million news articles by removing duplicates), we created a set of ~120K unique entities that are described by the 780 DBpedia categories and 225 concepts in MHDA-Kb. The figures below show two examples that illustrate entities spotted during entity extraction on a sample news article. A news article that has entities identified using this sequence of steps is selected for our study.

\begin{figure}[htbp]
\centerline{\includegraphics[width=80mm]{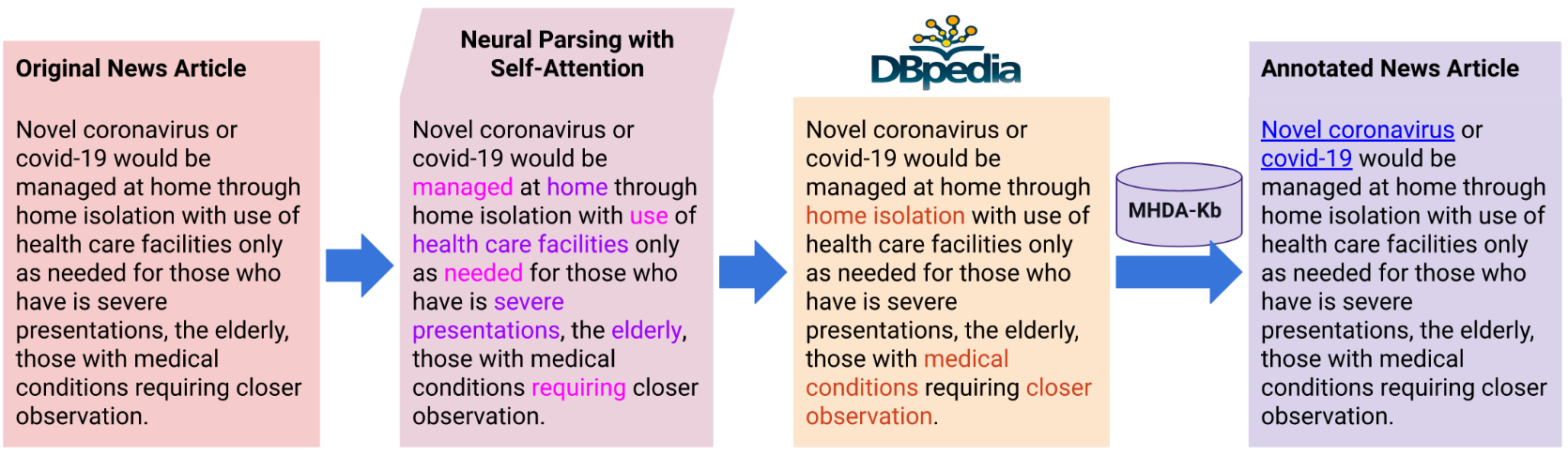}}

\vspace*{-0.3cm}
\caption{Example entity extraction-I using Semantic Filtering}
\label{fig}
\end{figure}

\begin{figure}[htbp]
\centerline{\includegraphics[width=80mm]{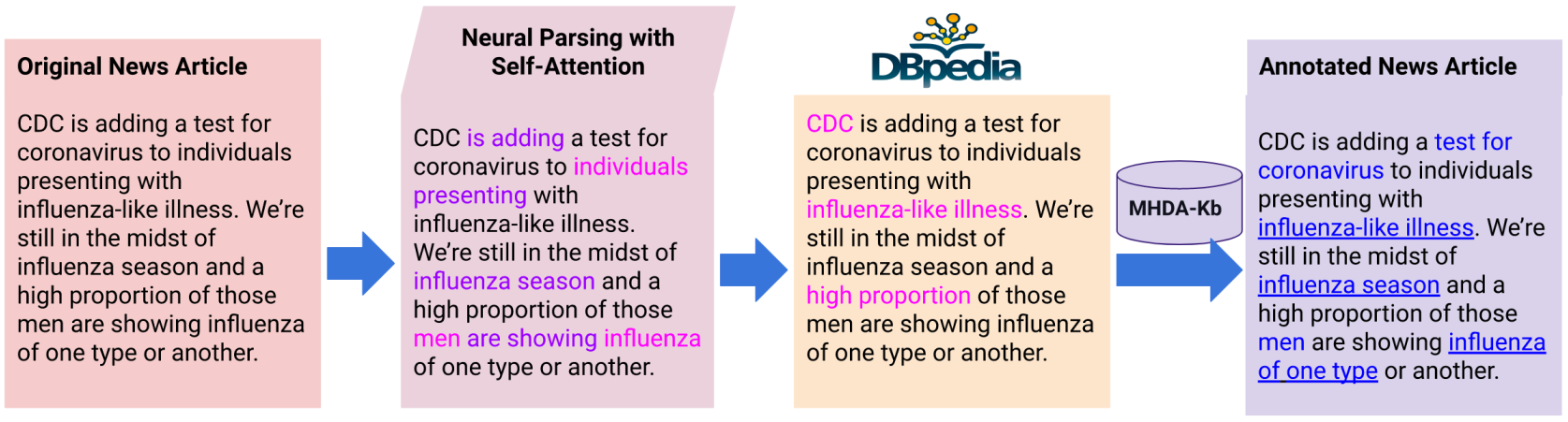}}

\vspace*{-0.3cm}
\caption{Example entity extraction-II using Semantic Filtering}
\label{fig}
\end{figure}

\section{Methods}
\label{sec:SM}

We propose to use three variations of the BERT model for representing news articles. In its basic form, we use vanilla BERT for encoding news articles. For the remaining two variations, we fine-tune BERT on a binary sequence classification task by independently training on two corpora  using masked language modeling (MLM) and next sentence prediction (NSP) objectives. The two corpora used are: 1) Subreddit Depression (\citeauthor{gkotsis2017characterisation}, \citeyear{gkotsis2017characterisation}; \citeauthor{gaur2018let}, \citeyear{gaur2018let}; \citeauthor{alambo2019question}, \citeyear{alambo2019question}); 2) A combination of subreddits: Crippling Alcoholism, Opiates, Opiates Recovery, and Addiction (abbreviated COOA), each consisting of Reddit posts about drug abuse. Subreddit Depression has 760049 posts across 121795 Redditors, and COOA has 1416765 posts from 46183 users, both consisting of posts from the years 2005 - 2016. Reddit posts belonging to subreddits depression or COOA are considered positive classes and the 380444 posts from control group ($\sim$10K subreddits unrelated to mental health) as negative classes. We use the following settings for training our BERT model for sequence classification: training batch size of 16, maximum sequence length of 256, Adam optimizer with learning rate of 2e-5, number of training epochs set to 10, and a warmup proportion of 0.1. We used 40\%-60\% split for training and testing sets for creating the BERT models and achieved a test accuracy of 89\% for Depression-BERT and 78\% for Drug Abuse-BERT. We set the size of the training set smaller than the testing set for generalizability of our models. In this manuscript, we refer to the BERT model fine tuned on subreddit depression as Depression-BERT or DPR-BERT, while the one fine tuned on subreddit COOA as Drug Abuse-BERT or DA-BERT.

In addition to using BERT for encoding news contents, we also use it for representing the entities in the background knowledge bases (i.e., PHQ-9, DAO, and DBpedia). Once we have encoded the news articles and the entities in the knowledge bases using vanilla BERT or fine-tuned BERT model, we generated depressiveness score, drug abusiveness score, and informativeness score corresponding to the entities in PHQ-9, DAO, and DBpedia respectively. The equation below gives the score of a news article for a category given one of the BERT models:

\begin{equation}
Score^{m}_{c}(news) = \frac{1}{|\mbox{E}_{KB}|}\sum^{|\mbox{E}_{KB}|}_{e=1} cossim~(\mbox{news}, e)
\label{eq:semantic_score}
\end{equation}
\\
\noindent where,\

$\mbox{m} \in \{\mbox{vanilla-BERT}, \mbox{DPR-BERT}, \mbox{DA-BERT}\}$


$\mbox{c} \in \{\mbox{informativeness}, \mbox{depressiveness}, \mbox{drug abuse\}}$

cossim~(\mbox{news}, \mbox{e}): cosine similarity between a news content
and an entity in KB

\mbox{KB} { - } a collection of entities present in PHQ-9, DBpedia, or DAO\\
\\
We used the base variant of the BERT model with 12 layers, 768 hidden units, and 12 attention heads. We use PyTorch 1.5.0+cu101 for fine-tuning our BERT models. All our programs were run on Google Colab’s NVIDIA Tesla P100 PCI-E GPU.

\section{Preliminary Results and Discussion}
\label{sec:EF}
In this section, we report the state-wise labels (i.e., depressive, drug abusive, informative) for each month obtained after summing the scores of news articles as described. The category with the highest cumulative score is set as the label for a state.

\begin{figure*}[!htbp]
  \begin{center}
    \includegraphics[width=120mm, scale=1.5, 
    trim=6.0cm 1.5cm 5.5cm 0.5cm]{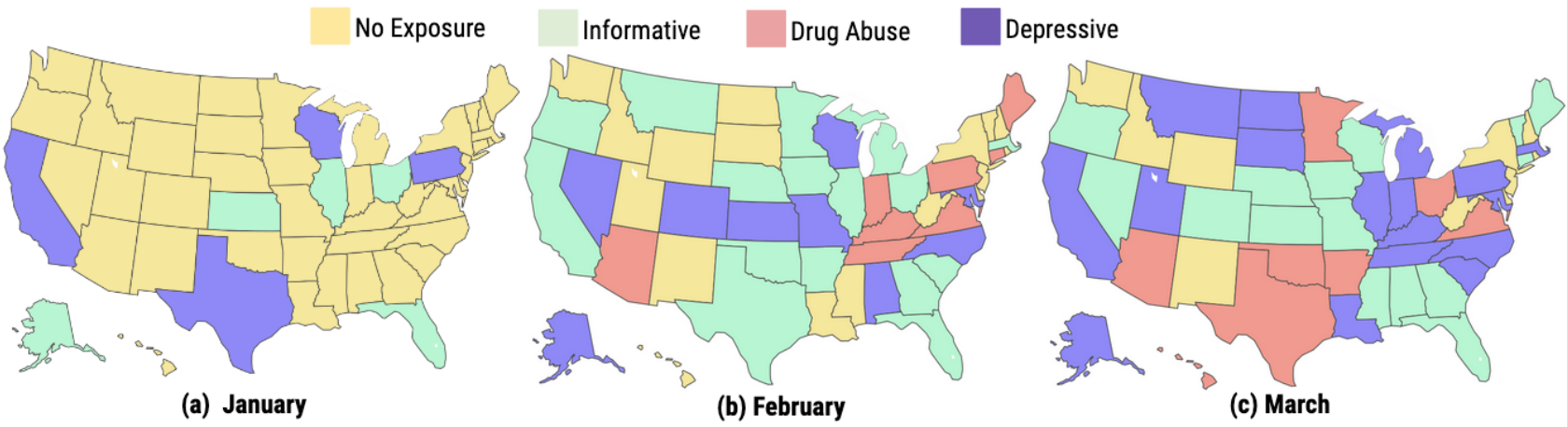}
  \end{center}
  \caption{vanilla BERT modeling of Depressiveness, Drug Abuse, and Informativeness in US states.}
  \label{fig:architecture}
\end{figure*}

Using vanilla-BERT (Figure 5), we can see that no state shows exposure to news content  on drug abuse in January. Going from February to March, we see depressive news content  move from inner-most states such as Missouri, Kansas, and Colorado to border states such as California, Montana, North Dakota, and Louisiana, making way for informative news content. Further, there are fewer states exposed to drug-related news content than those exposed to depressive or informative news content in February or March. Particularly, Arizona and Virginia show consistent exposure to drug-related news content in February and March.

\begin{figure*}[!htbp]
  \begin{center}
    \includegraphics[width=120mm, scale=1.5, 
    trim=6.0cm 1.5cm 5.5cm 0.5cm]{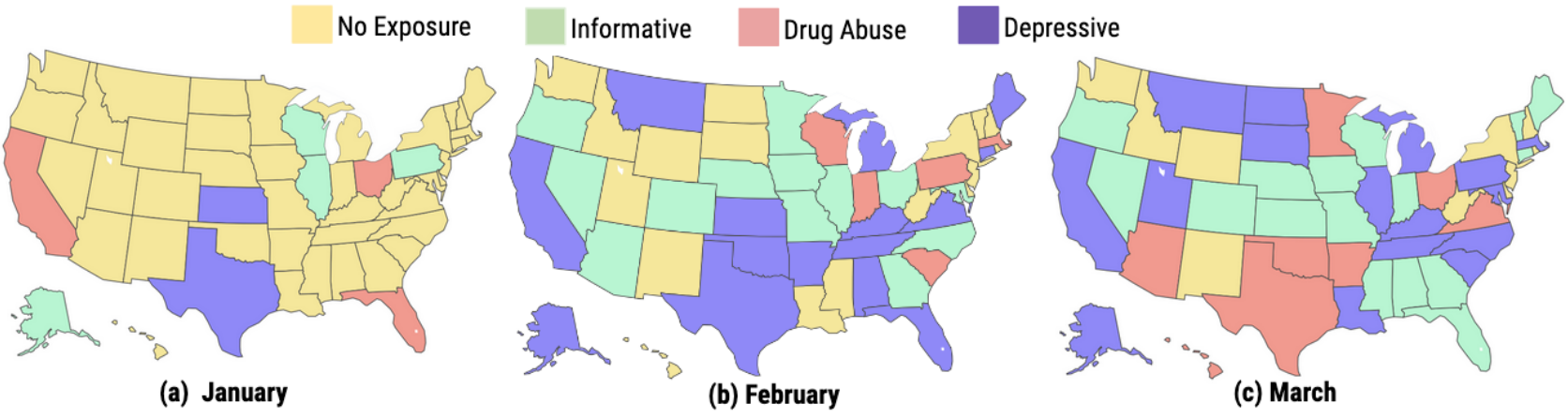}
  \end{center}
  \caption{Depression-BERT (DPR-BERT) modeling of Depressiveness, Drug Abuse, and Informativeness in US states}
  \label{fig:architecture}
\end{figure*}

Using depression-BERT, as shown in Figure 6, we see that states such as Texas, and Kansas are exposed to depressive news content for the month of January and February while states such as California, Montana, Alaska, and Michigan show higher consumption of depressive news content  in February and March. With regard to informativeness, we see an overall even distribution of informative news content  across the nation in February and March. Further, we see a few midwest states showing relatively higher instances of news content  that are informative than depressive in February and March. It’s interesting to see a few southern states such as Oklahoma, Texas, and Arkansas transition from exposure to depressive news content in the month of February to drug use related news content in the month of March.

\begin{figure*}[!htbp]
  \begin{center}
    \includegraphics[width=120mm, scale=1.5, 
    trim=6.0cm 1.5cm 5.5cm 0.5cm]{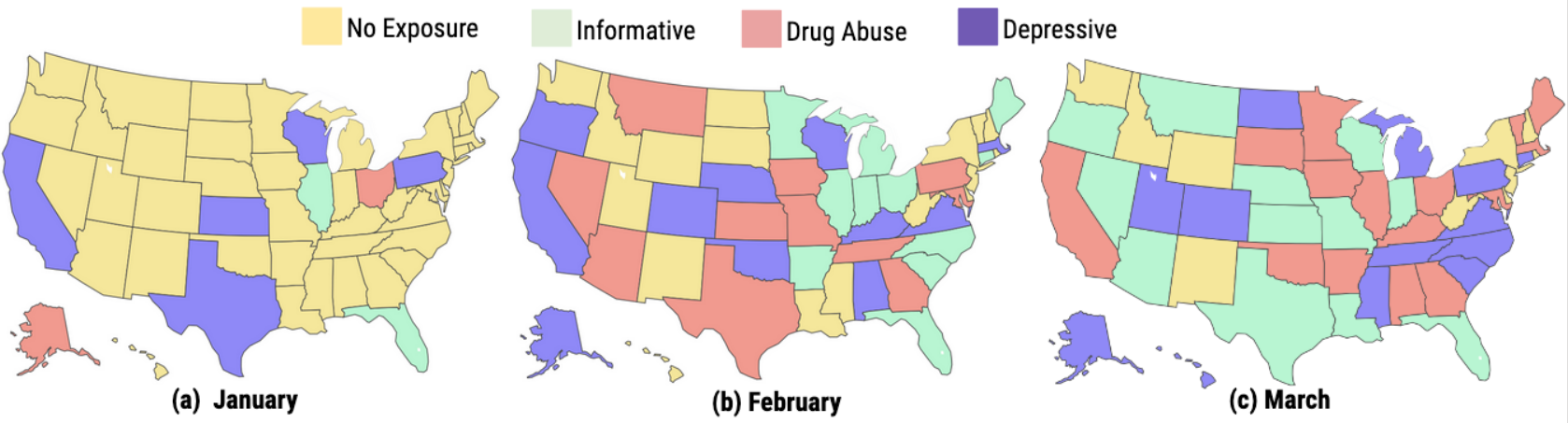}
  \end{center}
  \caption{Drug Abuse BERT (DA-BERT) modeling of Depressiveness, Drug Abuse, and Informativeness in US states}
  \label{fig:architecture}
\end{figure*}

Using Drug Abuse-BERT model (Figure 7), states such as Texas, and Wisconsin shift from exposure of depressive news content  in January to exposure of drug-related news content  in February, while states such as California, and Oklahoma transition from exposure to depressive news content  in February to drug-related news content  in March. Further, we see the informativeness of news content sweeping from the east to the midwest, to parts of the south, and to some parts of the west from February to March.  

Our results show that a fine-tuned BERT model cleanly separates the thematic categorical scores to a state. For instance, using DA-BERT  for the month of March, the drug abuse score for the state of California is much higher than the score of depressiveness or informativeness for the same state. However, with the vanilla BERT model, the three scores computed for the various states and months are marginally different. Moreover, the results using DPR-BERT or DA-BERT capture the state-level ranking of mental disorders by Mental Health America \footnote{https://www.mhanational.org/issues/ranking-states} better than vanilla-BERT; for a few states, the fine-tuned BERT models identify more months to have media exposure to depression or drug abuse news content. \\

\vspace{-0.5em}
\begin{table}[!htbp]
\centering
\begin{tabular}{p{2cm}p{1.8cm}p{1.8cm}p{1.6cm}} \toprule[2pt]
\textbf{MHA States with high ~~DPR and DA} & \textbf{vanilla-BERT
 (Months with depression/drug abuse)} & \textbf{DA-BERT 
(Months with depression/drug abuse)} & \textbf{DPR-BERT 
(Months with depression/drug abuse)} \\ \midrule[1.2pt]
Tennessee & Feb, Mar & Feb, Mar & Feb, Mar \\
Alabama & Feb & Feb, Mar & Feb \\
Oklahoma & Mar & Feb, Mar & Feb, Mar \\
Kansas & Feb & Jan, Feb & Jan, Feb \\
Montana & Mar & Feb & Feb, Mar \\
South Carolina & Mar & Mar & Feb, Mar \\
Alaska & Feb, Mar & Jan, Feb, Mar & Feb, Mar \\
Utah & Mar & Mar & Mar \\
Oregon & None & Feb & None \\
Nevada & Feb & Feb & None \\ \bottomrule[2pt]
\end{tabular}
\caption{Evaluation of base and domain-specific BERT models for MHA states over the period of three months (January, February, and March). These three months showed high dynamicity in COVID-19 spread.}
\label{tab:Red-stat}
\vspace{-2em}
\end{table}

As indicated in Table 1, we report months showing predominant media exposure to either depressive or drug abuse news articles using the three variants of BERT model. We use 10 of the 13 states recognized as showing high prevalence of mental disorders according to a report by Mental Health America on overall mental disorder ranking. The 3 states not included in this table are Washington, Wyoming, and Idaho. We did not consider these 3 states as these states were not in our dataset cohort. For the Mental Health America (MHA) report, we make a practical assumption that each of the three months is either depressive or drug abusive for each state. Thus, our objective is to maximize the number of months with exposure to depressive/drug abuse news content for each of the 10 states. We can see in Table 1 that fine-tuned BERT models help identify more months to having exposure to depressive or drug abuse news content than vanilla BERT does for the 10 states. For example, using DA-BERT, five states are identified to have at least two months showing exposure to depressive/drug abuse news content while DPR-BERT identifies six states to having been exposed to depressive/drug abuse news content for two months. On the other hand, vanilla-BERT identifies only two states with depressive/drug abuse news content for two months.  To compare models with one another and against the report by Mental Health America (MHA), we compute a Jaccard Index between each pair of models and each model against the report from MHA. The equation below computes Jaccard similarity between the results of two models or a model’s results with an MHA report.

\begin{equation}
J(m_1, m_2) = \sum_{i\ \in\ S} ^ {|S|}\frac{m_1^{M} \cap m_2^{M}}{m_1^{M} \cup m_2^{M}}\
\label{eq:jaccard_similarity}
\end{equation}

\noindent where,


\(m_1, m_2 \in {\mbox{\{vanilla-BERT, DPR-BERT, DA-BERT, MHA\}}}\)

$S \mbox{ - } \mbox{Set of States in the US (Table 1)}$

$m_1^{M}$, $m_2^{M}$: Number of depressive, drug abusive, or informative months for a state ``$\mbox{i}$''

We report inter-model and model-to-MHA Jaccard similarity scores computed using equation (2) in Figure 8.

\begin{figure}[!htbp]
\centerline{\includegraphics[width=80mm]{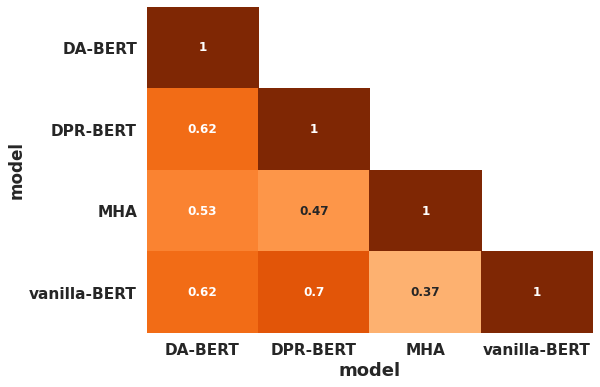}}

\vspace*{-0.3cm}
\caption{Inter-BERT model and BERT Model-to-MHA Jaccard Similarity Scores as a measure of closeness of model's prediction to an extensive survey on Mental Health America (MHA).}
\label{fig}
\end{figure}

As shown in Figure 8, DA-BERT gives the best results against MHA report in Jaccard similarity (0.53), which means DA-BERT identifies over half of the state-to-month instances in MHA. On the other hand, vanilla-BERT has a Jaccard similarity of 0.37 with MHA, which can be interpreted as vanilla-BERT identifies a little over one-third of the state-to-month instances in MHA. The best Jaccard similarity is achieved between DPR-BERT and vanilla-BERT (0.7); thus, 70\% of state-to-month mappings are shared between DPR-BERT and vanilla-BERT based on Jaccard index. It’s interesting to see DA-BERT has the same Jaccard similarity with vanilla-BERT and DPR-BERT, subsuming the former and being subsumed by the latter in terms of depressive/drug abusive months. 

\section{Conclusion}
\label{sec:C}
In this paper, we model  depressiveness, drug abusiveness, and informativeness of news articles to assess the dominant category characterizing each US state during each of the three months (Jan 2020 to Mar 2020).  We demonstrate the power of transfer learning by fine-tuning an attention-based deep learning model on a different domain and use the domain-tuned model for gleaning the nature of media exposure. Specifically, we use background knowledge bases for measuring depressiveness, drug abusiveness, and informativeness of news articles. We found out DA-BERT identifies the most number of state-to-month instances as being exposed to depressive or drug abuse news content according to the report from Mental Health America. In the future, we plan to incorporate background knowledge bases in our attention-based transfer learning framework to further investigate knowledge-infused learning (\citeauthor{kursuncu2019knowledge}, \citeyear{kursuncu2019knowledge}; \citeauthor{gaur2020knowledge}, \citeyear{gaur2020knowledge}).

\section*{ACKNOWLEDGEMENT}
\footnotesize
We acknowledge partial support from the National Institutes of Health (NIH) award MH105384-01A1: "Modeling Social Behavior for Healthcare Utilization in Depression". Any opinions, conclusions or recommendations expressed in this material are those of the authors and do not necessarily reflect the views of the  NIH.

\bibliographystyle{ACM-Reference-Format}
\bibliography{main} 

\end{document}